\documentclass[12pt,preprint]{aastex}
\def\degpoint{\ifmmode ^{\rm{o}}\!. \else $^{\rm{o}}\!.$\fi}

\newcommand{\ms}{\mbox{m\,s$^{-1}$}}

\newcommand{\Mjup}{\mbox{M$_{\rm Jup}$}}

\newcommand{\Mearth}{\mbox{M$_{\oplus}$}}

\newcommand{\ltsimeq}{\raisebox{-0.6ex}{$\,\stackrel
         {\raisebox{-.2ex}{$\textstyle <$}}{\sim}\,$}}

\begin{document}

\title{The Frequency of Low-Mass Exoplanets. III. Toward $\eta_{\oplus}$ 
at Short Periods }

\author{Robert A.~Wittenmyer\altaffilmark{1}, 
C.G.~Tinney\altaffilmark{1}, R.P.~Butler\altaffilmark{2}, Simon 
J.~O'Toole\altaffilmark{3}, H.R.A.~Jones\altaffilmark{4}, 
B.D.~Carter\altaffilmark{5}, J.~Bailey\altaffilmark{1}, 
J.~Horner\altaffilmark{1} }
\altaffiltext{1}{Department of Astrophysics, School of Physics, 
University of NSW, 2052, Australia}
\altaffiltext{2}{Department of Terrestrial Magnetism, Carnegie 
Institution of Washington, 5241 Broad Branch Road, NW, Washington, DC 
20015-1305, USA}
\altaffiltext{3}{Australian Astronomical Observatory, PO Box 296, Epping, 1710, 
Australia}
\altaffiltext{4}{Centre for Astrophysics Research, University of 
Hertfordshire, College Lane, Hatfield, Herts AL10 9AB, UK}
\altaffiltext{5}{Faculty of Sciences, University of Southern Queensland, 
Toowoomba, Queensland 4350, Australia}
\email{
rob@phys.unsw.edu.au}

\shorttitle{The Frequency of Low-Mass Exoplanets. III.}
\shortauthors{Wittenmyer et al.}

%-------------------------------------------------------------------
\begin{abstract}

\noindent Determining the occurrence rate of terrestrial-mass planets 
($m_p < 10$\Mearth) is a critically important step on the path towards 
determining the frequency of Earth-like planets ($\eta_{\oplus}$), and 
hence the uniqueness of our Solar system.  Current radial-velocity 
surveys, achieving precisions of 1\,\ms, are able to detect 
terrestrial-mass planets and provide meaningful estimates of their 
occurrence rate.  We present an analysis of 67 solar-type stars from the 
Anglo-Australian Planet Search specifically targeted for very 
high-precision observations.  When corrected for incompleteness, we find 
that the planet occurrence rate increases sharply with decreasing 
planetary mass.  Our results are consistent with those from other 
surveys: in periods shorter than 50 days, we find that 1.5\% of stars 
host a giant ($m_p > 100$\Mearth) planet, and that 18.5\% of stars host 
a planet with $m_p < 10$\Mearth.  The preponderance of low-mass planets 
in short-period orbits is in conflict with formation simulations in 
which the majority of terrestrial-mass planets reside at larger orbital 
distances.  This work gives a hint as to the size of $\eta_{\oplus}$, 
but to make meaningful predictions on the frequency of terrestrial 
planets in longer, potentially habitable orbits, low-mass terrestrial 
planet searches at periods of 100-200 days must be made an urgent 
priority for ground-based Doppler planet searches in the years ahead.

\end{abstract}

\keywords{planetary systems -- techniques: radial velocities }

%--------------------------------------------------------------------
\section{Introduction}

To date, 20 extrasolar planets are 
known\footnote{http://www.exoplanets.org} with minimum masses 
(m~sin~$i$) less than 10\Mearth.  Hundreds more planet candidates with 
sizes smaller than a few Earth radii, and therefore potentially 
terrestrial in nature, have been identified by the \textit{Kepler} 
spacecraft \citep{borucki11a, borucki11b}.  It is clear that 
terrestrial-mass planets exist, but what is not yet clear is the 
percentage of stars that form such planets, and how often those planets 
survive post-formation dynamical interactions in order to be observed 
today.  The frequency of Earth-mass planets in the habitable zone, often 
referred to as $\eta_{\oplus}$, is a key science driver for the 
\textit{Kepler} and \textit{CoRoT} missions.  Within this decade, these 
space missions are anticipated to provide an estimate of $\eta_{\oplus}$ 
with unparalleled accuracy and precision.  However, the bottleneck for 
large transit surveys has always been the radial-velocity follow-up to 
obtain mass estimates for planet candidates.  Until the multifarious 
candidates identified by these spacecraft have mass determinations, 
radial-velocity surveys capable of 1\,\ms\ precision will make a 
critical contribution to constraining $\eta_{\oplus}$ and the planetary 
mass function.  This work is prompted by the recent results of 
\citet{howard10}, who presented estimates for the occurrence rate of 
planets in short periods ($P<50$ days) from the NASA-UC Eta-Earth 
survey.  Our aim is to verify those results by using their methods on 
our own independent data set.

The Anglo-Australian Planet Search (AAPS) has undertaken two long, 
continous observing campaigns (48 and 47 nights) with the aim of 
detecting low-mass planets in periods shorter than 50 days 
\citep{monster}.  These two ``Rocky Planet Search'' campaigns have 
targeted a total of 54 bright, stable stars with spectral types between 
G0 and K5.  This strategy of observing through a dark lunation 
facilitates the detection of planets in that period regime by 
suppressing the window function near one lunar month (29 days).  
Previous AAPS planet discoveries arising from these observing campaigns 
include HD~16417b \citep{16417paper}, 61~Vir~b,c,d \citep{61vir}, and 
HD~102365b \citep{tinney11}.  In addition, we have chosen a subset of 
the AAPS main program stars for observation at high precision by 
requiring a signal-to-noise (S/N) of at least 300 per epoch.  Since the 
aim is a single-epoch radial-velocity precision of 1\,\ms\ (in the 
absence of stellar jitter), we designate these as ``One Meter Per 
Second'' (OMPS) stars.  There are 67 OMPS stars in the AAPS target list, 
of which the 54 Rocky Planet Search targets form a subset.  All of these 
stars receive at least 20 minutes of integration time per epoch, in 
order to average over the stellar p-mode oscillations \citep{otoole08}.

We have previously presented detailed simulations of planet 
detectability based on data from the 24 stars observed in the first 
Rocky Planet Search campaign in 2007: exploring the frequency of planets 
with periods less than $\sim$20 days in \citet{monster}, and 
investigating the nature of the ``period valley'' in 
\citet{foreverpaper}.  In this work, we consider the entirety of the 
OMPS and Rocky Planet Search targets, applying our simulation algorithms 
to the 67 stars in the AAPS sample which have data with the highest 
velocity precisions.  In particular, we seek to compute the completeness 
of this sample and to estimate the occurrence rate of low-mass planets 
with periods $P<50$ days.  We choose this period bin to match the 
primary focus of \citet{howard10} and compare our results with theirs.  
In Section 2, we present the input data sample and the methods used to 
obtain detection limits.  In Section 3, we compute the completeness of 
the sample and determine the occurrence rate of planets in four mass 
bins, before drawing conclusions in Section~4.

%--------------------------------------------------------------------
\section{Data Properties and Analysis Methods}

In this work, we focus on the 67 ``OMPS'' stars in the AAPS program 
which have the highest radial-velocity precision.  The OMPS target stars 
were selected on the basis of being apparently inactive (log 
$R^{'}_{HK}<-4.7$) and lacking a massive (i.e.~$>$10\Mjup) companion, 
and bright enough to obtain S/N$>$300 in no more than 30 minutes of 
integration time.  This represents essentially all of the AAPS target 
stars down to $V=6.50$, with a few additional stars being added between 
$V=6.5-7.0$ to fill in observing gaps in right ascension.

\begin{figure}
\plotone{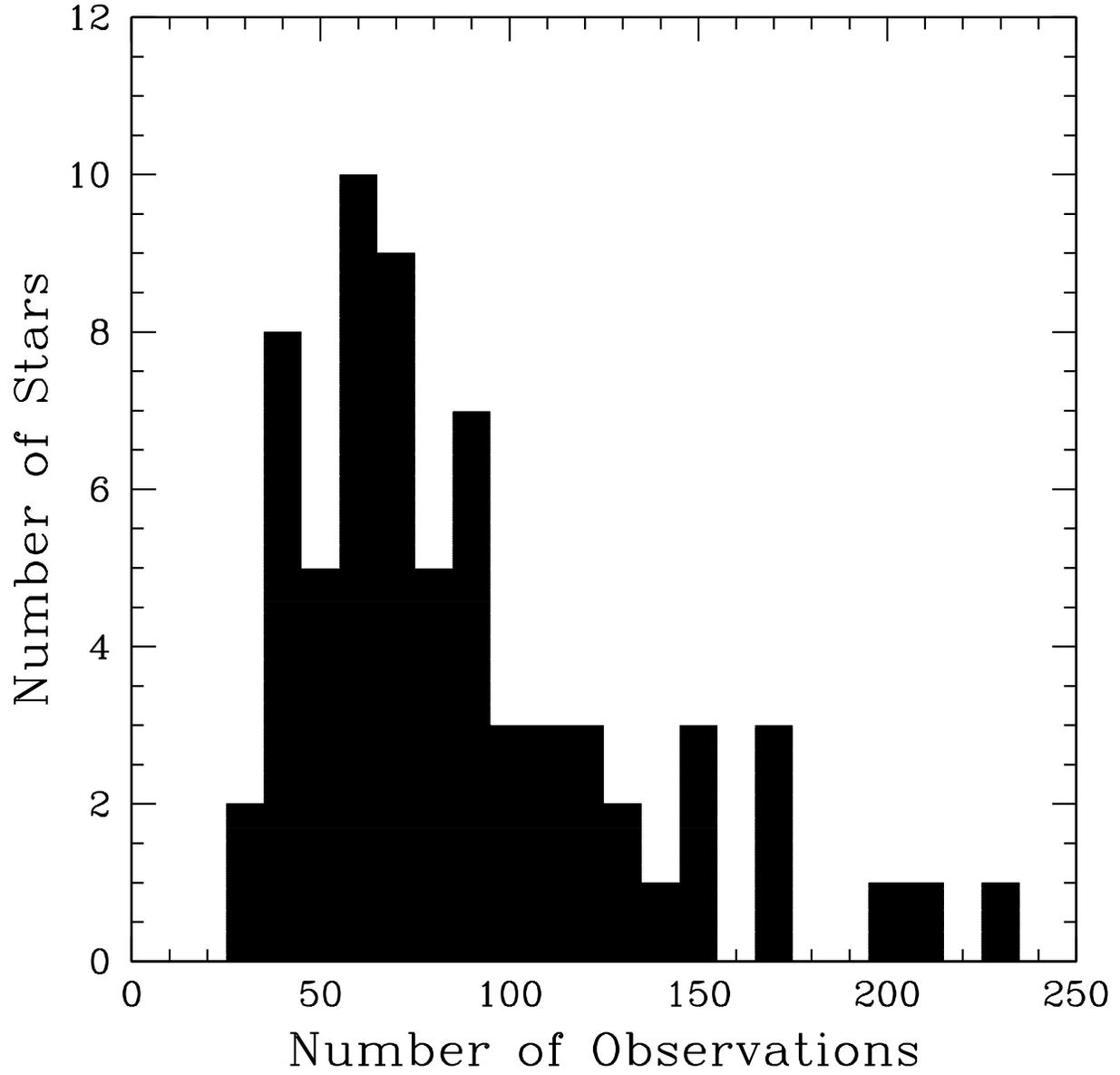}
\caption{Histogram of the number of observations for the 67 AAPS stars 
considered here.  Stars found to host low-mass planets contribute to the 
high-N tail. }
\label{datahisto}
\end{figure}

We fit for and removed velocity trends due to stellar companions, as 
well as the orbits of known planets.  The data are summarized in 
Table~\ref{rvdata1}, and a histogram of the number of observations is 
shown in Figure~\ref{datahisto}.  The parameters of all known 
planets in this sample are given in Table~\ref{planets}.

We determined the detectability of planets in these data by adding 
simulated Keplerian signals to the velocity data, then increasing the 
velocity amplitude ($K$) of the artificial planet until 100\% of signals 
at that period were recovered.  For a given $K$ at a given orbital 
period $P$, we use a grid of 30 values of periastron passage $T_0$.  A 
signal was considered recovered if its period in a standard Lomb-Scargle 
periodogram \citep{lomb76, scargle82} had a false-alarm probability 
(FAP) of less than 0.1\%.  Trials were also performed at recovery rates 
ranging from 10\% to 90\%.  The simulated planets had periods between 2 
and 1000 days, with 100 trial periods evenly spaced in log $P$.  As in 
\citet{howard10}, the simulated planets had zero eccentricity.  This 
method is identical to that used in our previous work (e.g.~Wittenmyer 
et al.~2006, 2009, 2010, 2011).

%--------------------------------------------------------------------
\section{Results and Discussion}

\begin{figure}
\plotone{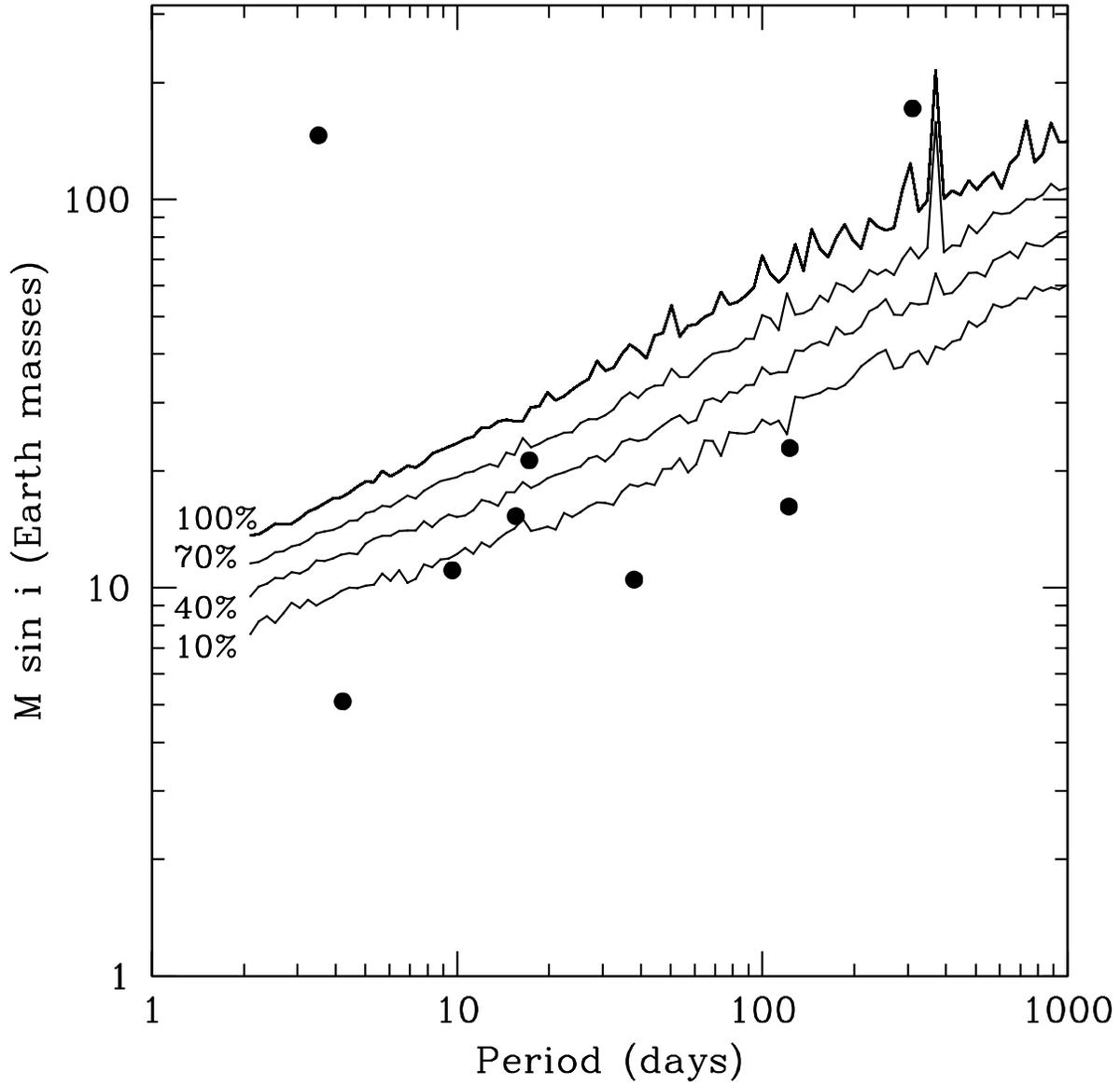}
\caption{Detection limits for planets in circular orbits, averaged over 
the 67 stars considerd here.  The contours indicate the fraction of 
injected planets which were recovered.  Filled circles represent 
detected planets in the sample; a further 9 planets with large masses or 
long periods are off the scale. }
\label{detect}
\end{figure}

Figure~\ref{detect} shows the mass limits (m~sin~$i$) averaged over all 
67 stars at four recovery rates: 100\%, 70\%, 40\% and 10\%.  Detected 
planets in the sample are represented by large filled circles.  
Throughout the discussion on the results of these simulations, we use 
``mass'' to refer to the projected planetary mass m~sin~$i$ obtained 
from radial-velocity measurements.  Since the inclination of the system 
is generally unknown, the planetary mass (m~sin~$i$) is a minimum value.

\subsection{Completeness Correction}

The 67 stars considered here host 18 currently known planets orbiting 12 
stars.  However, to determine the \textit{underlying} frequency of 
planets in the sample, we need to use the detectabilities we obtain from 
the simulations to correct these detections for our survey's varying 
completeness as a function of planet period and planet mass.  Moreover, 
because these detectabilities vary from star to star, we need to make 
this completeness correction on a star-by-star, rather than on a 
whole-of-survey basis.

We therefore estimate how many planets have been ``missed'' from our 
survey as a whole, by calculating the ``missed planet'' contribution for 
each detected planet using

%For a detection of a planet with period $P$ and mass $M$, the simulation 
%results give a recovery fraction $f_R(P,M)$ for each star.  For each 
%detected planet, then, the number of ``missed planets'' is given by

\begin{equation}
N_{missed} = \Bigg[\frac{1}{N_{stars}}\sum_{j=1}^{N_{stars}} 
f_{R,j}(P_i,M_i)\Bigg]^{-1}-1,
\end{equation}

\noindent where $f_{R,j}(P_i,M_i)$ is the recovery fraction as a 
function of mass $M_i$ at period $P_i$ (for the $i$th detected planet), 
and $N_{stars}$ is the total number of stars in the sample ($N=67$).  
There are $i$ detected planets in the sample, and $j$ stars in total.  
For a detected planet with period $P_i$ and mass $M_i$, each star 
contributes a detectability $f_R(P_i,M_i)$ between 0 and 1 to the sum in 
Equation~(1).  The quantity $f_R(P_i,M_i)$ is the fraction of simulated 
planets with period $P_i$ and mass $M_i$ which were recovered.  In this 
way, we compute the detectability averaged over the whole sample for 
each detected planet at the specific ($P_i,M_i$) of that planet.  This 
approach, also employed in \citet{jupiters}, thus accounts for the 
non-uniformity of detectability across the sample.  We show the results 
of these calculations in the column labeled ``Method~1'' of 
Table~\ref{missedplanets}.

This method for estimating the number of ``missed planets'' (i.e.~the 
correction for survey completeness) is nearly identical to that used by 
\citet{howard10}, except that they defined ``completeness'' as the 
fraction of \textit{stars} for which a planet of mass $M$ at period $P$ 
was recovered in 100\% of trials.  That is, each star contributes either 
0 or 1 to the sum in Equation~(1). Since that work considered only the 
100\% recovery level, a star whose detection limit falls just short of 
the mass for a given planet would be counted as \textit{never} able to 
detect that planet, whereas the true detectability may still be 
significant (i.e.~$>$90\%).  We have used our simulation results (at 
100\% recovery) to estimate the number of ``missed planets'' using this 
method, and this is given in Table~\ref{missedplanets} as ``Method~2.'' 
The completeness computations of \citet{howard10} excluded stars with 
detected planets; our results for Method~2 thus excluded the 12 planet 
hosts, leaving 55 stars.  The last column of Table~\ref{missedplanets} 
(``Method~3'') gives the results obtained using this method when we 
include the 12 planet-host stars in the calculations.

We see a pronounced difference in the results obtained by Methods 1 and 
2/3 for the lowest-mass planets: when the completeness is a binary 
function (either a planet is detected 100\% of the time or it is never 
detected), the number of missed planets is poorly sampled.  Indeed, this 
can lead to nonsensical results, e.g.~an infinite number of missed 
planets.  This unphysical result occurs when considering the detection 
of planets with properties that match those of the super-Earth 
HD~115617b (=61~Vir~b; Vogt et al.~2010).  At the mass and period of 
HD~115617b, none of the data sets for the 55 non-planet-hosting stars in 
our sample enabled the detection of the simulated planet in 100\% of 
trials.  This resulted in $f_R(P,M)=0$, and hence an infinite number of 
missed planets by Equation~(1).  HD~115617 is an unusual target in that 
we have a large number of observations ($N=139$) and it is a very stable 
star, with a residual velocity rms (to the three-planet fit) of only 
2.3\,\ms.  This extreme example highlights two important points to 
consider in the estimation of the frequency of extremely low-mass 
planets: first, that meaningless results are obtained when the 
completeness approaches zero, and second, inhomogeneities in 
planet-search data sets require a detailed, star-by-star approach to 
best determine the true underlying frequency of low-mass planets.  Great 
caution is therefore required when interpreting results from 
survey-completeness simulations such as these, especially when 
considering terrestrial-mass planets, where current radial-velocity 
surveys are heavily affected by incompleteness.

For the AAPS data considered here, the missed-planet correction used by 
\citet{howard10} (given in Table~\ref{missedplanets} as Method~2), is 
clearly not useful.  Even when planet-hosting stars are included 
(Method~3), the correction for missed planets gives a result that is 
unjustifiably overestimated.  This is due to the uneven data density for 
our sample, as the detection limits achievable depend heavily on the 
number of observations \citep{jupiters}.  The AAPS sample has a mean 
$N_{obs}=88\pm$45, whereas the Keck sample has a mean $N_{obs}=40\pm$22.

Nonetheless, the high-$N_{obs}$ tail seen in Figure~\ref{datahisto} 
reflects a reality for all radial-velocity programs: stars with 
candidate low-mass planets (e.g.~61~Vir) are prioritised and receive a 
larger number of observations.  This directly leads to the situation 
described above, where a very low-mass planet \textit{only} has a high 
detectability for that one star, and can have detectabilities of, 
e.g.~only 10\% for the remaining targets in the sample.  Because the 
Keck sample of \citet{howard10} has a somewhat more uniform distribution 
in $N_{obs}$, their data are less prone to this feature, and their 
method is therefore less problematic.  It is, however, inappropriate for 
our data and we adopt Method~1 for this work and all subsequent 
discussion.

\subsection{Close-in Planets ($P<50$ days)}

To directly compare our results with those of \citet{howard10}, we now 
focus on periods shorter than 50 days.  While \citet{howard10} used mass 
bins of width log$_{10}(\Delta\,M_{Earth})=0.5$, our sample has fewer 
detected planets, so we use four mass bins of width 
log$_{10}(\Delta\,M_{Earth})=1.0$.

\begin{figure}
\plotone{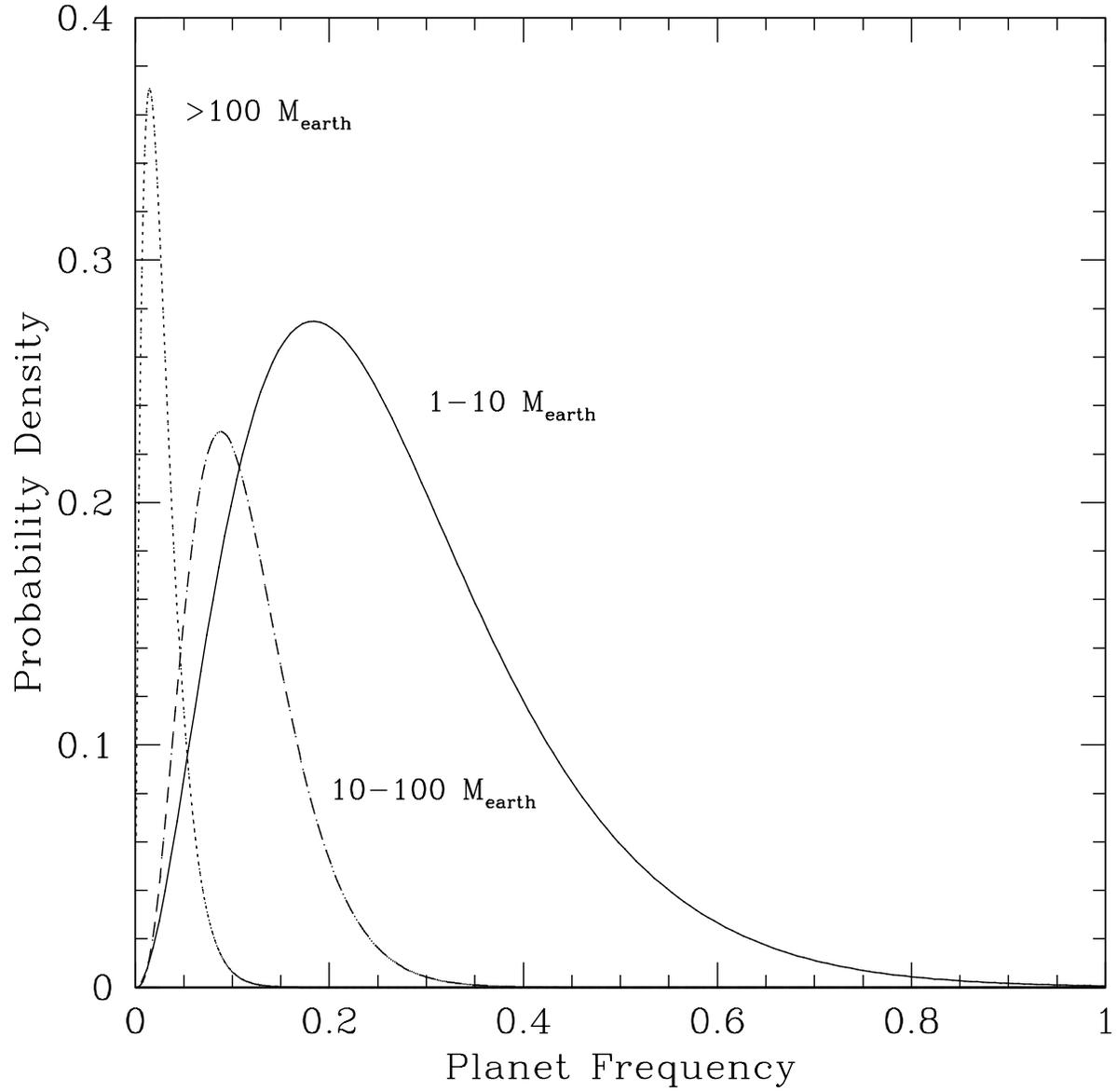}
\caption{Binomial probability density functions computed for three mass 
bins and periods less than 50 days.  Less-massive planets are clearly 
more prevalent. }
\label{probs}
\end{figure}

\begin{figure}
\plotone{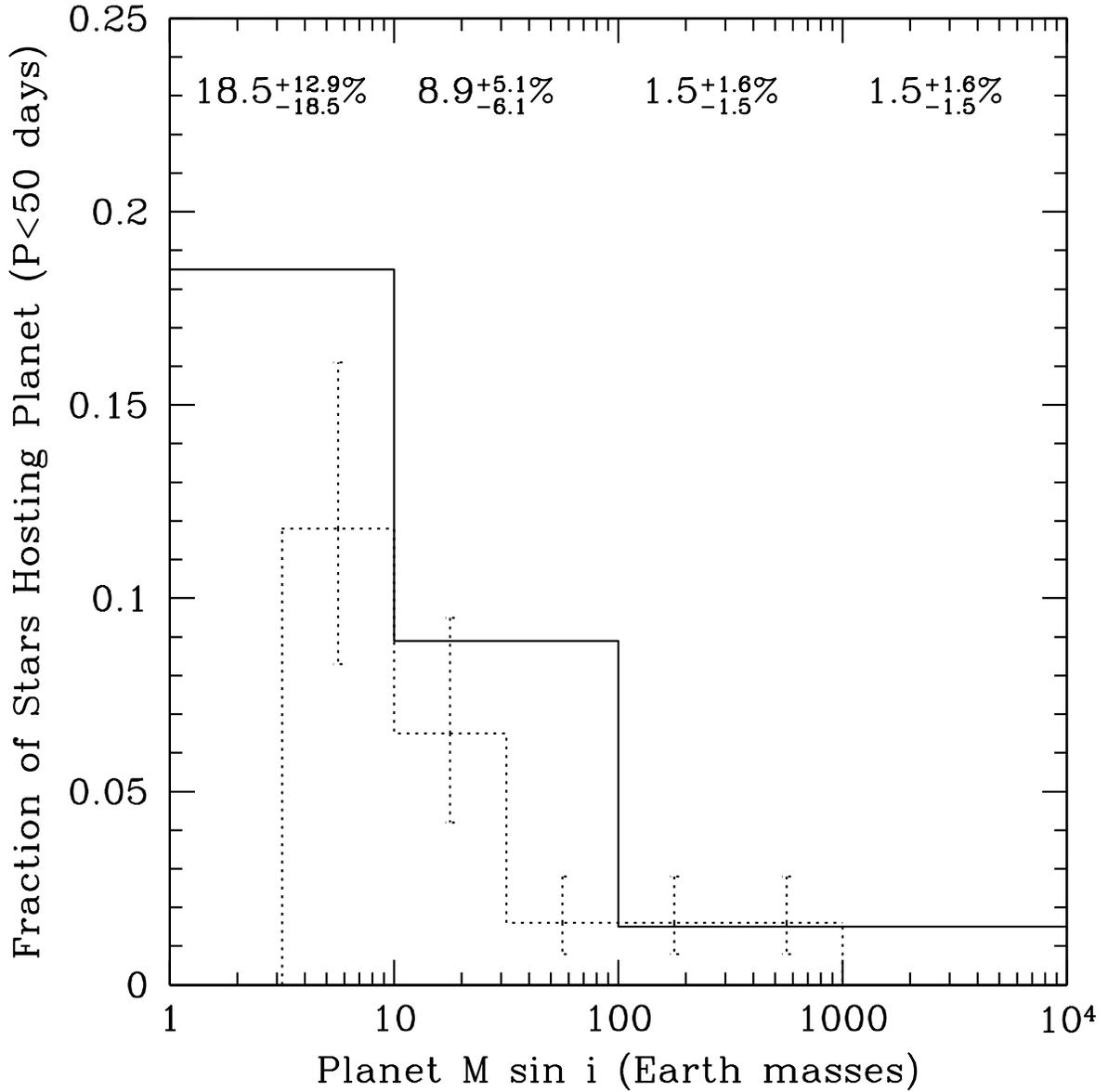}
\caption{Planet frequency as a function of mass, from this work (solid 
histogram) and compared with \citet{howard10} (dashed histogram).  The 
two sets of results are consistent within their uncertainties. }
\label{compare}
\end{figure}

We estimate the frequency of planets in each bin using binomial 
statistics, after \citet{howard10}.  That is, we compute the binomial 
probability of detecting exactly $k$ planets in a sample of $n$ stars, 
with the underlying probability $p$ of hosting a planet.  We compute 
this over all $p$ to find the most probable value (Figure~\ref{probs}).  
In this way, we estimate the planet frequency and its 1-sigma 
uncertainty (68.3\% confidence interval) for each of the four mass bins.  
The planet frequencies obtained are then adjusted for incompleteness by 
multiplying each bin's frequency and its uncertainty by a factor 
$(N_{detected}+N_{missed})/N_{detected}$.  The results are shown in 
Table~\ref{frequency}.  The uncertainties on our measured planet 
frequency are large, owing to the small number of detections (7 planets 
with $P<$50 days) compared to the Keck survey (16 planets).  As previous 
studies have shown (e.g.~Howard et al.~2010, Wittenmyer et al.~2010, 
O'Toole et al.~2009c), planet frequency increases as planet mass 
decreases; more low-mass planets are found despite the fact that they 
are much more difficult to detect.  Figure~\ref{compare} plots the 
derived planet frequencies from this work and those of \citet{howard10} 
for direct comparison.  Our results are consistent with those of the 
NASA-UC Eta-Earth survey: we find that $18.5^{+12.9}_{-18.5}$\% of stars 
host terrestrial-mass planets ($M_p<10$\Mearth) at periods of less than 
50 days.

%--------------------------------------------------------------------
\section{Conclusions}

Our data are consistent with the estimation of \citet{mayor09} that 
30$\pm$10\% of solar-type stars host a planet with m~sin~$i$\ltsimeq 
30\Mearth\ and $P<50$ days.  These results, and those of other 
radial-velocity planet search teams, support the idea that close-in 
terrestrial-mass planets (or ``super-Earths'' with $m_p < 10$\Mearth) 
are quite common in orbital periods less than 50 days.  These 
observational data are, at present, in disagreement with 
planet-formation simulations (e.g.~Mordasini et al~2009, Ida \& Lin 
2005, 2008, Kornet \& Wolf 2006) which predict an under-abundance of 
such planets orbiting inside of $\sim$1~AU.  \citet{idalin08} instead 
predict a large number of super-Earths to accumulate near the ice line, 
beyond 2~AU.  Such objects are completely undetectable by current 
radial-velocity surveys, but the observational data in hand suggest that 
the planet population synthesis models require significant revision in 
order to reproduce the high abundance of close-in super-Earths for which 
there is now a growing body of evidence.

The frequency of habitable Earth-like planets ($\eta_{\oplus}$) is a key 
quantity to measure as we seek to understand the frequency of habitable 
environments in the Universe.  However, it is important to note that 
while these results (and those of Howard et al.~2010) provide hints on 
the size of $\eta_{\oplus}$, they do not determine $\eta_{\oplus}$ 
directly.  Given their orbital periods ($P<$50d), and therefore 
semi-major axes (0.24--0.30~AU), none of the terrestrial-mass planets 
probed by these studies are actually habitable -- they are all far too 
hot.

A key next step for this research will be extending searches for the 
lowest-mass planets to larger orbital periods (and so semi-major axes).  
If we can at least understand the trends in the frequency with which 
planet formation makes planets as a function of period, at periods from 
50 to 100 and even 150d, then we will be in a much better position to 
make {\em robust} predictions as to the frequency with which habitable 
terrestrial planets (i.e.~planets in 200-400d orbits) are formed around 
solar-type stars.  Low-mass terrestrial planet searches at 100-200d must 
be made an urgent priority for ground-based Doppler planet searches in 
the years ahead \citep{guedes08, endl09}.

%--------------------------------------------------------------------
\acknowledgements

We gratefully acknowledge the UK and Australian government support of 
the Anglo-Australian Telescope through their PPARC, STFC and DIISR 
funding; STFC grant PP/C000552/1, ARC Grant DP0774000 and travel support 
from the Australian Astronomical Observatory.  RW is supported by a UNSW 
Vice-Chancellor's Fellowship.

This research has made use of the Exoplanet Orbit Database and the 
Exoplanet Data Explorer at exoplanets.org.  We have also made use of 
NASA's Astrophysics Data System (ADS), and the SIMBAD database, operated 
at CDS, Strasbourg, France.

%--------------------------------------------------------------------

%----------------------------------------------------------
% This table updated 8 Feb, includes all data.

\begin{deluxetable}{lrrr}
\tabletypesize{\scriptsize}
\tablecolumns{4}
\tablewidth{0pt}
\tablecaption{Summary of Radial-Velocity Data }
\tablehead{
\colhead{Star} & \colhead{$N$} & \colhead{RMS} & \colhead{$<\sigma>$} \\
\colhead{} & \colhead{} & \colhead{(\ms)} & \colhead{(\ms)}
 }
\startdata
\label{rvdata1}
HD 142 & 74 & 10.97 & 3.19 \\
HD 1581 & 97 & 3.59 & 1.26 \\
HD 2151 & 175 & 4.28 & 0.84 \\
HD 3823 & 70 & 5.82 & 1.75 \\
HD 4308 & 107 & 4.51 & 1.36 \\
HD 7570 & 43 & 6.34 & 1.53 \\
HD 10360 & 61 & 4.48 & 1.33 \\
HD 10361 & 60 & 4.56 & 1.23 \\ 
HD 10700 & 231 & 3.68 & 1.09 \\ 
HD 13445 & 60 & 4.87 & 1.93 \\
HD 16417 & 113 & 3.99 & 2.53 \\ 
HD 20794 & 134 & 3.45 & 1.07 \\ 
HD 20807 & 89 & 4.48 & 1.50 \\
HD 23249 & 79 & 3.47 & 0.62 \\ 
HD 26965 & 94 & 4.79 & 0.83 \\ 
HD 27442 & 87 & 7.11 & 0.87 \\
HD 28255A & 61 & 7.23 & 1.69 \\ 
HD 38382 & 36 & 4.83 & 1.69 \\
HD 39091 & 59 & 5.57 & 2.23 \\
HD 43834 & 123 & 4.98 & 1.15 \\ 
HD 44120 & 32 & 3.55 & 1.71 \\
HD 45701 & 30 & 5.86 & 2.00 \\
HD 53705 & 125 & 4.55 & 1.60 \\ 
HD 53706 & 38 & 3.02 & 1.47 \\
HD 65907A & 58 & 6.20 & 1.75 \\
HD 72673 & 55 & 3.66 & 1.25 \\ 
HD 73121 & 38 & 5.86 & 1.92 \\
HD 73524 & 78 & 5.25 & 1.63 \\ 
HD 75289 & 41 & 5.80 & 1.77 \\
HD 84117 & 123 & 5.45 & 1.70 \\ 
HD 100623 & 75 & 5.01 & 1.09 \\ 
HD 102365 & 153 & 2.76 & 1.11 \\ 
HD 102438 & 47 & 4.59 & 1.71 \\
HD 108309 & 55 & 3.54 & 1.25 \\
HD 114613 & 198 & 5.68 & 0.98 \\ 
HD 115617 & 139 & 2.32 & 1.96 \\ 
HD 122862 & 93 & 4.22 & 1.68 \\ 
HD 125072 & 68 & 5.28 & 1.19 \\
HD 128620 & 99 & 4.06 & 0.93 \\ 
HD 128621 & 134 & 3.58 & 0.70 \\ 
HD 134060 & 86 & 5.65 & 1.44 \\
HD 134987 & 67 & 2.96 & 1.30 \\
HD 136352 & 146 & 4.74 & 1.27 \\ 
HD 140901 & 102 & 10.36 & 1.26 \\
HD 146233 & 62 & 5.25 & 1.16 \\
HD 156274B & 92 & 4.83 & 1.29 \\
HD 160691 & 167 & 2.25 & 0.89 \\
HD 168871 & 62 & 4.91 & 1.92 \\
HD 172051 & 49 & 3.37 & 1.13 \\
HD 177565 & 90 & 3.98 & 1.15 \\
HD 189567 & 79 & 5.55 & 1.63 \\
HD 190248 & 208 & 4.05 & 0.96 \\
HD 191408 & 168 & 4.20 & 1.17 \\
HD 192310 & 146 & 3.93 & 1.15 \\
HD 193307 & 76 & 4.27 & 1.79 \\
HD 194640 & 70 & 4.83 & 1.46 \\ 
HD 196761 & 38 & 4.78 & 1.01 \\
HD 199288 & 68 & 5.48 & 2.23 \\
HD 207129 & 114 & 4.95 & 1.22 \\
HD 210918 & 65 & 5.32 & 1.28 \\
HD 211998 & 40 & 14.69 & 3.02 \\
HD 212168 & 42 & 5.59 & 1.67 \\
HD 214953 & 76 & 4.98 & 1.73 \\
HD 216435 & 74 & 7.05 & 2.08 \\
HD 216437 & 49 & 4.92 & 1.74 \\
HD 219077 & 60 & 3.89 & 1.36 \\
HD 221420 & 70 & 4.77 & 1.51 \\
\enddata
\end{deluxetable}

%----------------------------------------------------------

%----------------------------------------------------------
\begin{deluxetable}{lr@{$\pm$}lr@{$\pm$}lr@{$\pm$}ll}
% \rotate
\tabletypesize{\scriptsize}
\tablecolumns{5}
\tablewidth{0pt}
\tablecaption{Planets From This Sample }
\tablehead{
\colhead{Planet} & \multicolumn{2}{c}{Period} & \multicolumn{2}{c}{M sin $i$ } &
\multicolumn{2}{c}{$a$ } & \colhead{Discovery Ref.} \\
\colhead{} & \multicolumn{2}{c}{(days)} &
\multicolumn{2}{c}{(\Mearth)} & \multicolumn{2}{c}{(AU)} & \colhead{}
 }
\startdata
\label{planets}
HD 142 b & 350.4 & 1.5 & 419.5 & 38.9 & 1.05 & 0.02 & \citet{tinney02} \\
HD 4308 b & 15.609 & 0.007 & 13.0 & 1.4 & 0.118 & 0.009 & \citet{udry06} \\
HD 13445 b & 15.7656 & 0.0005 & 1280.8 & 69.9 & 0.114 & 0.002 & \citet{queloz00} \\
HD 16417 b & 17.24 & 0.01 & 22.1 & 2.0 & 0.14 & 0.01 & \citet{16417paper} \\
HD 27442 b & 430.8 & 0.8 & 508.5 & 29.5 & 1.27 & 0.02 & \citet{butler01} \\
HD 75289 b & 3.50918 & 0.00003 & 146.3 & 6.8 & 0.048 & 0.001 & \citet{udry00} \\
HD 102365 b & 122.1 & 0.3 & 16.0 & 2.6 & 0.46 & 0.04 & \citet{tinney11} \\
HD 115617 b & 4.2150 & 0.0006 & 5.1 & 0.5 & 0.050201 & 0.000005 & \citet{61vir} \\
HD 115617 c & 38.021 & 0.034 & 18.2 & 1.1 & 0.2175 & 0.0001 & \citet{61vir} \\
HD 115617 d & 123.01 & 0.55 & 22.9 & 2.6 & 0.476 & 0.001 & \citet{61vir} \\
HD 134987 b & 258.19 & 0.07 & 505.3 & 6.4 & 0.81 & 0.02 & \citet{vogt00} \\
HD 134987 c & 5000 & 400 & 260.6 & 9.5 & 5.8 & 0.5 & \citet{jones10} \\
HD 160691 b & 644.9 & 0.6 & 534.4 & 18.8 & 1.53 & 0.02 & \citet{butler01} \\
HD 160691 c & 4060 & 49 & 641.0 & 32.3 & 5.2 & 0.1 & \citet{mcc04} \\
HD 160691 d & 9.641 & 0.002 & 9.1 & 1.0 & 0.093 & 0.001 & \citet{santos04} \\
HD 160691 e & 308.7 & 0.7 & 156.4 & 12.4 & 0.94 & 0.01 & \citet{pepe07} \\
HD 216435 b & 1332 & 14 & 405.6 & 33.4 & 2.59 & 0.05 & \citet{jones03} \\
HD 216437 b & 1354 & 6 & 714.5 & 34.9 & 2.54 & 0.04 & \citet{jones02} \\
\enddata
\end{deluxetable}
%----------------------------------------------------------

%----------------------------------------------------------

\begin{deluxetable}{llll}
\tabletypesize{\scriptsize}
\tablecolumns{4}
\tablewidth{0pt}
\tablecaption{Missed Planets in the Sample }
\tablehead{
\colhead{Planet} & \colhead{Method 1} & \colhead{Method 
2\tablenotemark{a}} & \colhead{Method 3\tablenotemark{b}}
}
\startdata
\label{missedplanets}
HD 142 b & 0.0 & 0.1 & 0.1 \\
HD 4308 b & 1.4 & 6.9 & 5.1 \\
HD 13445 b & 0.0 & 0.0 & 0.0 \\
HD 16417 b & 0.3 & 1.0 & 1.0 \\
HD 27442 b & 0.0 & 0.0 & 0.0 \\
HD 75289 b & 0.0 & 0.0 & 0.0 \\
HD 102365 b & 4.7 & 54.0 & 21.3 \\
HD 115617 b & 7.0 & Inf & 21.3 \\
HD 115617 c & 1.2 & 12.8 & 7.4 \\
HD 115617 d & 2.1 & 54.0 & 21.3 \\
HD 134987 b & 0.0 & 0.1 & 0.0 \\
HD 134987 c & \nodata\tablenotemark{c} & \nodata & \nodata \\
HD 160691 b & 0.0 & 0.1 & 0.0 \\
HD 160691 c & \nodata & \nodata & \nodata \\
HD 160691 d & 3.3 & 17.3 & 10.2 \\
HD 160691 e & 0.0 & 0.1 & 0.1 \\
HD 216435 b & \nodata & \nodata & \nodata \\
HD 216437 b & \nodata & \nodata & \nodata \\
\enddata

\tablenotetext{a}{After Howard et al.~(2010)}
\tablenotetext{b}{Same as Howard et al.~(2010) but including 
detectabilities from the planet hosts also.}
\tablenotetext{c}{The simulations here considered only periods shorter 
than 1000 days, so detectability information is not available for these 
four long-period planets.}
\end{deluxetable}
%----------------------------------------------------------

%----------------------------------------------------------

\begin{deluxetable}{llll}
\tabletypesize{\scriptsize}
\tablecolumns{4}
\tablewidth{0pt}
\tablecaption{Short-Period Planet Frequencies }
\tablehead{
\colhead{Mass Bin} & \colhead{Detections} & \colhead{$N_{missed}$} & 
\colhead{Frequency}
}
\startdata
\label{frequency}
1-10\Mearth  & 2 & 10.3 & 18.5$^{+12.9}_{-18.5}$\% \\
10-100\Mearth  & 3 & 2.9 & 8.9$^{+5.1}_{-6.1}$\% \\
100-1000\Mearth  & 1 & 0.0 & 1.5$^{+1.6}_{-1.5}$\% \\
1000-10000\Mearth  & 1 & 0.0 & 1.5$^{+1.6}_{-1.5}$\% \\
\enddata
\end{deluxetable}
%----------------------------------------------------------

%----------------------------------------------------------

\end{document}